# High-Resolution Dynamic Full-Field Optical Coherence Microscopy: Illuminating Intracellular Activity in Deep Tissue

ERIKAS TARVYDAS,[1] AUSTĖJA TREČIOKAITĖ,[1] AND EGIDIJUS AUKSORIUS[1,*]

[1]*Center for Physical Sciences and Technology (FTMC), Vilnius, 10257, Lithuania*
*\*egidijus.auksorius@ftmc.lt*

**Abstract:** Dynamic full-field optical coherence microscopy (*d*-FF-OCM) is a label-free imaging technique that captures intrinsic subcellular motions to generate functional contrast. This dynamic approach yields images with fluorescence-like contrast, highlighting active structures without the need for fluorescent labels. However, current *d*-FF-OCM implementations struggle to image deep within highly scattering tissues at high resolution. Here, we present a new high-resolution *d*-FF-OCM system that overcomes these limitations, enabling much deeper high-resolution imaging in such tissues. The setup uses 100× oil-immersion objectives (NA 1.25) and a high-brightness, laser-pumped incoherent white light source to achieve nanometer-scale resolution at depths up to ~100 μm in highly scattering samples. We also incorporate real-time reference arm adjustment to maintain signal strength and contrast as the focus moves deeper into the sample. Using this system, we imaged fresh *ex vivo* mouse liver and small intestine with unprecedented depth and detail. In these tissues, the dynamic contrast clearly revealed fine structures not visible with conventional OCT—for example, the sinusoidal microvasculature and organized cell layers in the liver, as well as neural plexuses and crypts in the intestine—all visualized label-free. By bridging the gap between high-resolution and deep imaging in highly scattering tissue, this advance provides a powerful new tool for biological microscopy, with potential applications from fundamental research to rapid intraoperative pathology.

## 1. Introduction

Achieving high-resolution, label-free imaging deep within highly scattering biological tissues remains a major challenge in optical microscopy[1,2], critical for understanding fundamental biological processes and advancing clinical diagnostics. Optical Coherence Tomography (OCT) is a powerful imaging technique that has continuously evolved to achieve higher resolution and imaging depth[3]. While conventional scanning OCT modalities, including micro-OCT (*μ*OCT), which achieves ~1–2 μm resolution[4], offer deep tissue penetration, they inherently face a trade-off between lateral resolution and depth of focus, limiting their utility for high-speed and high-resolution volumetric acquisition[5]. Full-Field Optical Coherence Microscopy (FF-OCM) decouples lateral resolution from depth of focus by axial (*z*) scanning and camera-based parallel detection of *xy* planes enabling much higher resolution[6-9], albeit at the expense of slower volumetric acquisition due to the need for axial scanning. Despite the slow imaging speed of FF-OCM, in vivo imaging of the eye has been successfully demonstrated[10]. Fourier-domain FF-OCM advances imaging speed[11], and in some cases surpasses that of scanning OCT, but it remains constrained by the same resolution-depth trade-off and exhibits reduced sensitivity. Nevertheless, its ability to computationally correct for optical aberrations enables superior resolution for *in vivo* retinal imaging compared to standard scanning OCT[11,12]. Recently, pupil-scanning in Fourier-domain FF-OCM enabled 3D imaging on an *ex vivo* opaque cornea at a resolution of 0.29 μm and 0.5 μm in the lateral and axial planes, respectively[13]. However, such approaches remain to be validated in highly scattering tissues (e.g. liver), and their ~1 s per volume rate, while remarkable for volumetric imaging, may limit its suitability for applications where rapid 2D image acquisition is required, such as dynamic OCT. Similarly, other emerging techniques, like variants of quantitative phase imaging techniques[14] have demonstrated nanoscale topographic mapping and refractive index reconstruction in cornea. However, these methods primarily focus on structural and optical property quantification rather than intrinsic cellular dynamics, and their application to highly scattering tissues like liver, particularly for rapid dynamic imaging, remains to be fully explored. Beyond its spatial resolution, OCT images suffer from contrast that is often nonspecific, making it difficult to distinguish different tissue components. To overcome this limitation, a variety of functional contrast techniques—such as OCT angiography[15], elastography[16] and dynamic OCT[17]—have been developed. The latter is the simplest, as it does not require additional elements, but provides a great contrast improvement in live tissue by using temporal fluctuations in the OCT signal caused by intrinsic subcellular dynamics, such as metabolic processes, cytoskeletal movements, and organelle activity[17]. These fluctuations generate a fluorescence-like contrast, suppressing signals from static, highly scattering structures and highlighting active cellular components[18]. It has been elegantly demonstrated in dynamic Full-Field Optical Coherence Microscopy ($d$-FF-OCM)[19] by simply switching the piezo performing the phase-shifting *off* and relying on the subcellular tissue motion to generate the signal. This dynamic contrast enables $d$-FF-OCM to resolve cellular morphology, distinguish cell types, and visualize processes at high resolution, such as mitosis, rosette formation, cell motility, and intracellular dynamics[7,8,19]. Beyond biological research, $d$-FF-OCM has shown promise in biomedical applications, including rapid intraoperative cancer diagnostics[20-23]. Despite these capabilities, $d$-FF-OCM, like all optical microscopy techniques, can benefit from improvements in spatial resolution and imaging depth to enable the observation of smaller subcellular structures in tissue. To this end, some studies have employed NA ≈ 0.8 lenses in $d$-FF-OCM[6,7,24]. However, advancing to even higher numerical apertures (e.g., NA = 1.05 for transparent retinal explants[24]) presents challenges, primarily dispersion mismatch and insufficient light intensity at the sample, both of which reduce SNR. When two highly corrected, high-NA objective lenses are employed in a Linnik interferometer, particularly with a broadband light source like white light, chromatic dispersion mismatch reduces the contrast of the interference fringes, which in turn reduces OCT signal and thus SNR. Illuminating a large field of view (FOV) through high-NA objectives with LEDs, which are used for its large spatial and temporal incoherence, is challenging. LEDs have high *étendue* and low radiance, making it difficult to efficiently fill the objective's pupil and deliver sufficient light intensity across the FOV. Conversely, laser-based broadband sources, with their high radiance, overcome this *étendue* mismatch, enabling bright and uniform illumination for high-resolution, wide-field imaging. Although high–spatial-coherence lasers perform well in scanning OCT[4,25], in FF-OCM they produce strong speckle artifacts that degrade image quality—especially at greater depths—thereby limiting both penetration depth and SNR. In Fourier-Domain FF-OCM[26] and even in scanning OCT[27] coherent noise can be reduced by partially decreasing spatial coherence or employing other means[28]. However, in FF-OCM, spatial coherence should be reduced as much as possible to maximize coherence-noise suppression—while still uniformly filling the objective's pupil and covering the full FOV. In this study, we employ a bright, incoherent laser-pumped white light source that provides ample intensity to the sample through a 100× objective (NA = 1.25). This enables deep tissue high resolution imaging, free from coherent artifacts. Imaging depth is extended via real-time reference arm adjustment[29]. We illustrate its superior performance on *ex vivo* mouse tissue, such as liver and small intestine.

## 2. Methods

**Optical System.** The dynamic FF-OCM system, illustrated in Figure 1, is comprised of three primary components: a high-brightness incoherent white light source (LS–WL1, Lightsource.tech, Germany), a Linnik interferometer with matched high-numerical-aperture objectives (Plan N 100×, NA = 1.25, Olympus, Japan), and a high full-well-capacity (FWC) camera (Q–2HFW–CXP, Adimec, Netherlands). The light source is a bright, high-power and compact unit that generates white light via the fluorescent conversion of focused 450 nm laser light on a specialized converter phosphor. A maximum of 300 mW was coupled into a multimode fiber (1 mm core, NA = 0.39) and delivered to the system, where after collimation with lens L1 ($f$ = 25 mm), the light is spectrally filtered to remove the strong 450 nm laser peak using a dichroic beamsplitter (DMLP490R, Thorlabs) with a 490 nm cut-on wavelength. Lens L2 ($f$ = 150 mm) forms the image of the fiber tip onto the pupil planes of both 100× objective lenses. The resulting 6× magnification of the fiber tip onto the pupil planes of the objective lenses ensures proper filling, which is crucial for achieving optimal resolution. The light is then distributed between the sample and reference arms by a 50/50 beamsplitter. Up to 50 mW could be sent onto the sample through the objective. The sample was mounted on a translation stage equipped with a stepper motor (Zaber T-NA08A25-S) for axial scanning. The entire reference arm was mounted on a separate motorized translation stage, enabling automatic optical path length adjustment. This adjustment keeps the temporal and confocal gates aligned at greater depths, counteracting the effect of the refractive-index mismatch between tissue and immersion oil[29], and thus maintaining imaging up to >100 μm depth. The reference mirror was mounted on a piezo stack (Piezomechanik STr-25) to enable controlled phase-shifting for deriving static (conventional) FF-OCM images. The reference arm was attenuated to 10% (in the double pass) by inserting an OD 0.5 neutral density (ND) filter to optimize the SNR. An additional ND filter of the same thickness, but with 0 OD, was inserted into the sample arm to compensate for chromatic dispersion. The beamsplitter then recombined light from the sample and reference arms, and a tube lens L3, $f$ = 180 mm (SWTLU-C, Olympus) focused the resulting interferogram onto the camera.

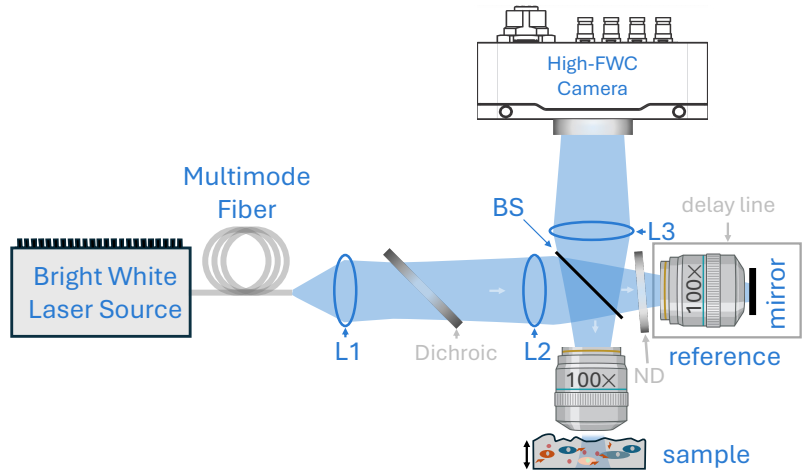

Fig. 1. Schematic of the dynamic FF-OCM system featuring two 100× oil immersion objectives in the interferometer and a bright laser-driven incoherent white light source. The light is delivered to the microscope by a multimode fiber and sent to an interferometer composed of a 50/50 beamsplitter and the objectives. Interferograms are detected by a high full well capacity (FWC) camera.

**Image Acquisition and Processing.** The camera acquired 1440 × 1440 images at 500 fps, with each pixel optimized for 2 million electrons full well capacity detection. The imaging field-of-view (FOV) was 173 μm × 173 μm with pixel size of 120 nm in the sample plane. Data transfer from the camera to the computer was managed by a BitFlow Cyton-CXP4 PCIe CoaXPress frame grabber, capable of transferring data at up to 25 Gb/s. Instrument control and synchronization were implemented via a National Instruments NI-PCIe6363 DAQ card. Data acquisition and analysis was managed by a custom LabVIEW application. Static FF-OCM images were derived by calculating a square root of $(I_1 - I_3)^2 + (I_2 - I_4)^2$, where $I_1, I_2, I_3$ and $I_4$ are images acquired at *0°, 90°, 180°* and *270°* phase changes imparted by different positions of the piezo in the reference arm. Dynamic FF-OCM images were acquired by deactivating the piezo-driven phase-shifting and instead relying on intrinsic subcellular movements within the tissue. These movements induce temporal intensity fluctuations on the camera, occurring on a millisecond timescale. Figure 2 illustrates how the raw image sequence is processed to generate the dynamic images. A 512-frame stack is acquired (Fig. 2a), and each frame is normalized to correct for camera's nonlinear response. For each pixel, its 512-point intensity time series is then analyzed in two ways: (i) computing the temporal standard deviation (STD) and (ii) performing a Fast Fourier Transform (FFT). STD analysis yields a single value per pixel, generally reflecting fluctuation strength. FFT analysis, conversely, provides a frequency spectrum for each pixel (Fig. 2b). For visualization, we here arbitrarily integrate each pixel's fluctuation spectrum into three bands—low (1–2 Hz), mid (2–100 Hz), and high (100–250 Hz)—which we map to blue, green, and red channels, respectively. These boundaries were chosen to emphasize slow, intermediate, and fast subcellular motions, but they can be adjusted to suit different samples or analysis goals. The integrated values were log-transformed and were assembled into a composite RGB image (Fig. 2c). Before the RGB image generation, to enhance overall contrast, the highest 1% of pixel values in R, G and B channels were saturated to full intensity, and, conversely, the lowest 1% of the values in those channels were set to zero intensity. For volumetric RGB rendering (Fig. 2d), we repeated this process over 100–250 z-steps, each taking ≈1 s to acquire and ≈5 s to save.

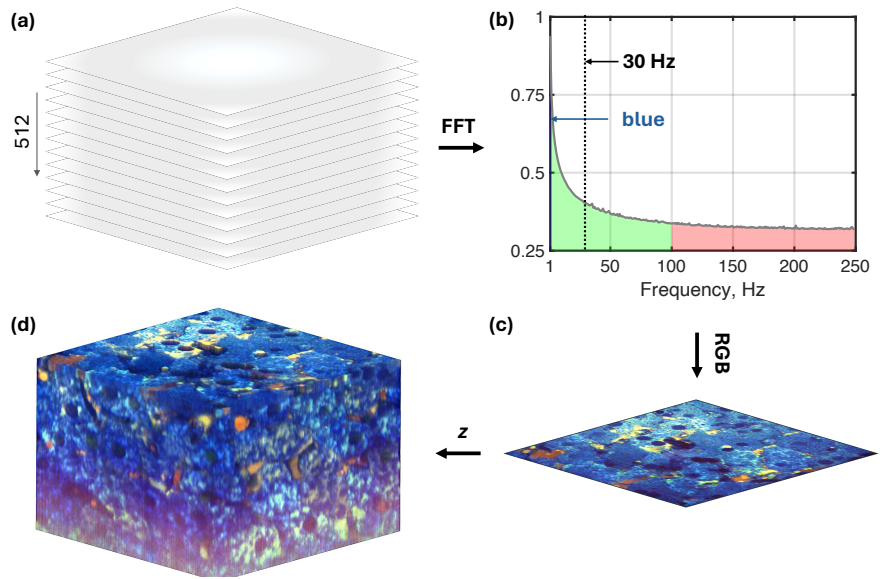

Fig. 2. Generation of dynamic FF-OCM images: a stack of 512 raw images (a) is Fast-Fourier Transformed (FFT) pixel-by-pixel to generate frequency spectrum (b) for each pixel. The spectrum in (b) represents the average over all pixel spectra from the image in (c). The spectrum in (b) is segmented into low (blue: 1–2 Hz), mid (green: 2–100 Hz), and high (red: 100–250 Hz) frequency bands. The blue (1–2 Hz) band is faint—see the blue arrow. Vertical line shows the upper frequency cut (of 30 Hz) that were used in most of the previous studies; (c) presents an RGB image with each pixel colored according to its relative spectral content; (d) shows one projection of 3D volumetric rendering (see Supplementary Video 1 for full video) resulting from the RGB images generated at multiple z positions, spanning 100 μm.

To correct the mismatch between the confocal and temporal gates, we first determined how much the entire reference arm must be shifted for each micrometer of sample movement. This was done by moving the sample 30 μm and adjusting the motorized translation stage in the reference arm to find the position that gave the highest image contrast. From this, we calculated the required reference arm shift per micrometer of sample displacement, assuming a homogeneous axial refractive index within the sample.

**Biological Samples**. For imaging, we used freshly excised organs of C57BL/6 mice. The mice were obtained from the local Life Sciences Center colonies. Animal studies were conducted in accordance with the requirements of the Directive 2010/63/EU and were approved by the Lithuanian State Food and Veterinary Service (permit No. G2-92). All mice were bred and kept at the animal facility of the Life Sciences Center of Vilnius University. Tissue was kept in PBS and imaged by slightly pressing the tissue surface against the 100 μm-thick microscopy coverslip glass plate to which the objective was coupled using immersion oil.

## 3. Results

**Resolution:** We have evaluated lateral and axial resolutions to be ~270 nm and ~500 nm, respectively, as shown in Fig. S1.

**Liver Imaging:** The fresh liver sample was cut in half and put on a coverslip with its cut side facing the coverslip. We then imaged it using both static and dynamic acquisition modes and two contrast-generation algorithms (STD and FFT) to extract the signal. Figure 3a shows a static (conventional) *en face* FF-OCM image acquired in ~1 s that matches the acquisition time used for dynamic FF-OCM acquisition of 512 raw images for direct comparison. *En face* (Fig. 3a) and axial (Fig. 3b) images reveal the dense packing of hepatocytes (20–30 μm diameter) with barely visible boundaries in between and offer little subcellular detail beyond dark nuclear voids. By contrast, *d*-FF-OCM processed via the temporal standard deviation (STD) method (Fig. 3c-d) produces a boost in contrast: cell borders and intracellular structures become clearly visible; filamentous networks of low-frequency fluctuations—likely related to mitochondrial dynamics[19]—emerge throughout many hepatocytes, alongside nuclear activity. Applying an FFT-based spectral segmentation to the same time series yields RGB-coded images (Fig. 3e-f) that further differentiate tissue components: hepatocyte bodies map predominantly to low-range frequencies, whereas sinusoids—capillary-like vessels found in the liver – map to a broader and higher range of frequencies. Sinusoids exchange substances between blood and hepatocytes[30] and are lined by endothelium and populated by erythrocytes, immune cells and other substances, like triglyceride-rich lipoproteins (chylomicrons) and plasma.

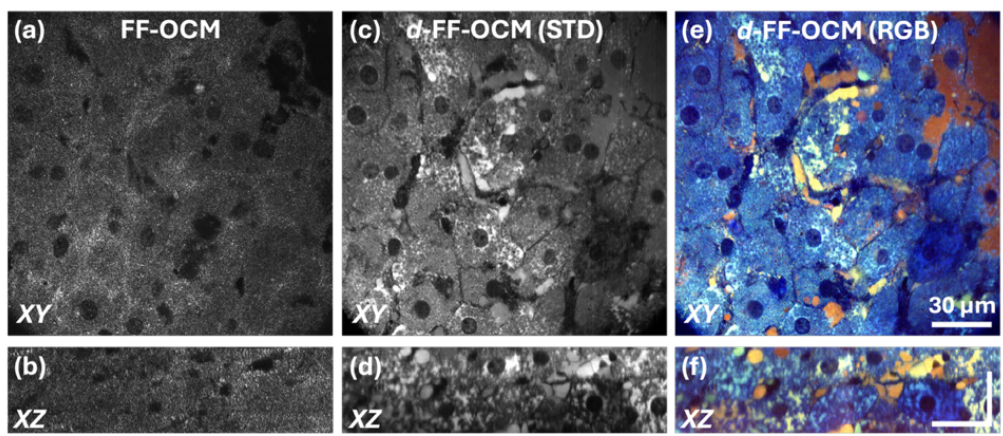

Fig. 3. Dynamic FF-OCM shows superior contrast over static FF-OCM. High-resolution *en face* (*XY*) and axial (*XZ*) views of fresh mouse liver acquired with conventional FF-OCM (a, b) versus dynamic FF-OCM (c–f). Dynamic imaging (c–f) enhances visualization of hepatocyte plates and cell-filled sinusoids. In the dynamic panels, blue denotes low-frequency (1–2 Hz), green mid-range (3–100 Hz), and red high-frequency (100–250 Hz) activity. Scale bar: 30 μm.

Dynamic FF-OCM images in RGB of deeper tissue layers (20–40 µm below the surface) of the same specimen are shown in Fig. 4a-c. It shows continuous cords of hepatocytes interlaced with sinusoids throughout this depth range (See Supplementary Video 2 for the full RGB stack). Four-times-magnified insets (Fig. 4d–f) highlight blood-filled sinusoids with discernible erythrocytes (Fig. 4d), a binucleated hepatocyte exhibiting spatially heterogeneous cytoplasmic and nuclear dynamics (Fig. 4e), and sinusoids containing cells consistent with Kupffer or NK cells based on size/morphology what could be platelets and immune cells (Fig. 4f), such as Kupffer, NK or T cells, for example. The two smallest structures in sinusoid, seen in Fig. 4f, appearing in light and dark red, are $1.65 \times 3.3$ µm and $2.5 \times 4.2$ µm in size, respectively. Importantly, high-contrast dynamic imaging persists even down to 100 µm, as demonstrated in the axial (XZ) views of Supplementary Fig. S3e–f; for example, the 80 µm slice in S3d still resolves hepatocyte and sinusoidal activity across a broad frequency spectrum. Moreover, we chose to use 0-250 Hz range instead of the typical 0-30 Hz because of better contrast, as shown in Supplementary Fig. S4.

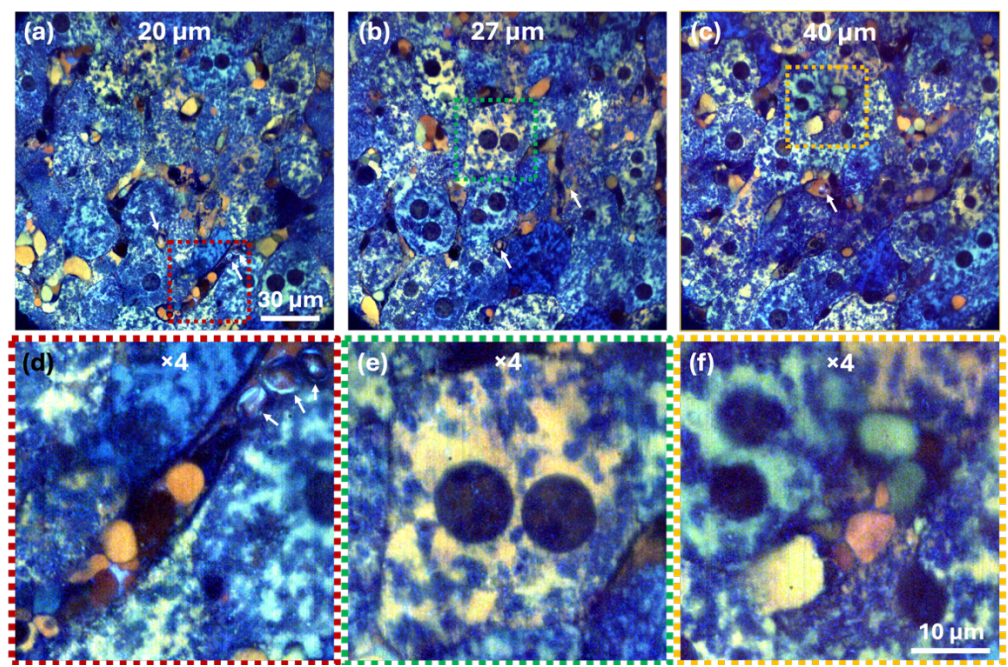

Fig. 4. Imaging subcellular dynamics across depth in fresh *ex vivo* mouse liver with dynamic FF-OCM. Images, taken from Supplementary Video 2, and correspond to the images recorded at depths of 20 µm (a), 27 µm (b), and 40 µm (c), clearly showing hepatocytes and sinusoids throughout the tissue. Panel (d) is a 4× magnified view of the red box in (a), highlighting a blood-filled sinusoid with visible erythrocytes (white arrows) and other cells. Panel (e) is a 4× magnified view of the red box in (b), showing a binucleated hepatocyte whose cytoplasm exhibits heterogeneous dynamic activity and whose nuclei also display motion. Panel (f) is a 4× magnified view of the red box in (c), featuring a sinusoid containing small blood elements—likely platelets and immune cells. In all the panels, color encodes subcellular dynamics: blue (1–2 Hz), green (3–100 Hz), and red (100–250 Hz).

To demonstrate the versatility of this method, we imaged the mouse small intestine from both the serosal and mucosal sides (Figures 5 and 6, respectively).
**Small Intestine Imaging (mucosal side):** Figure 5 presents images of two mouse gut samples imaged from the mucosal side, highlighting the villi (see Supplementary Videos 3 and 4 for full depth imaging). Epithelial cells, predominantly enterocytes, are organized in a mosaic pattern at the villus apex, as depicted in Figures 5a and 5d.

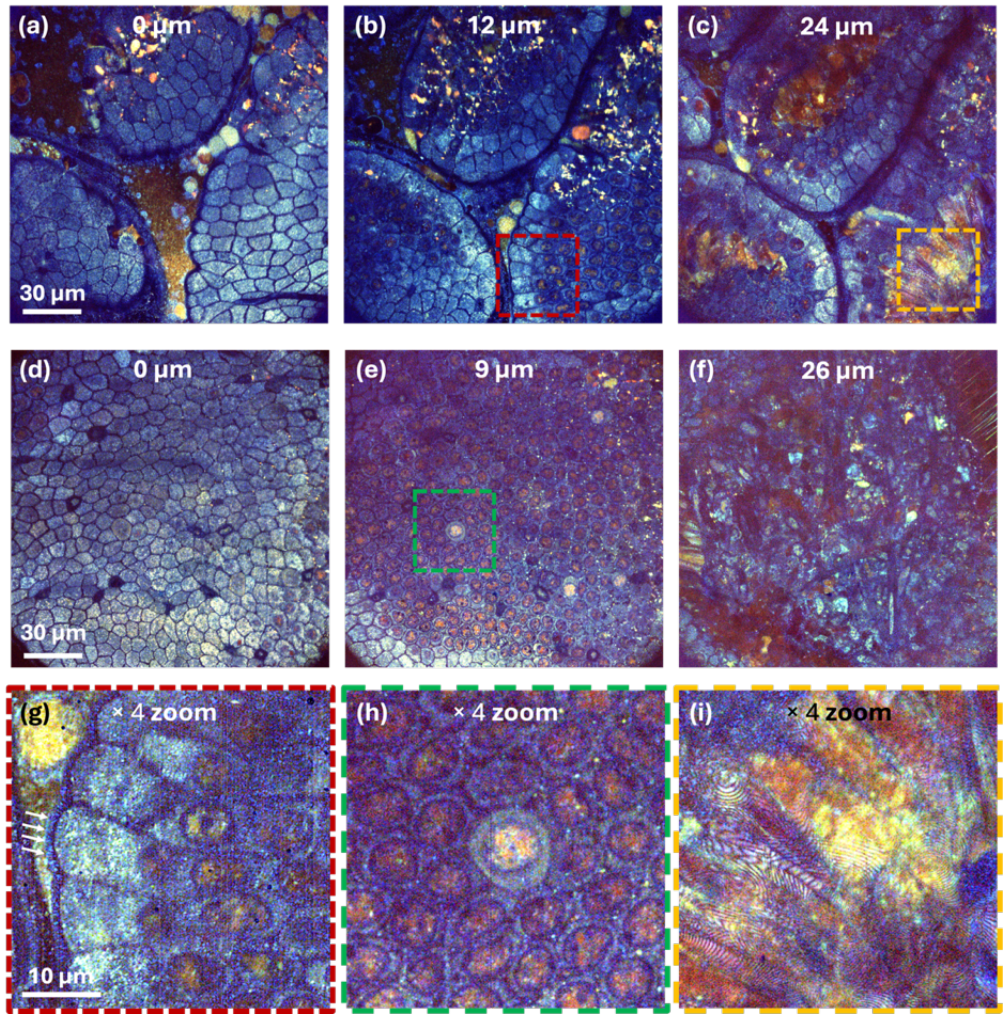

Fig. 5. Dynamic FF-OCM reveals depth-dependent subcellular architecture and activity in *ex vivo* mouse intestinal mucosa. *En face* images at three increasing depths are shown for sample #1 (a–c) and sample #2 (d–f), taken from Supplementary Videos 3 and 4, illustrating the epithelial layer overlying the lamina propria. Panels g–i present 4× magnified detail views of key features from these depth series: (g) epithelial microvilli (white arrows) and nuclei; (h) a putative goblet cell among enterocytes; (i) mesenchymal elements within the lamina propria. Dynamic fluctuations are color-coded by frequency: blue (1–2 Hz), green (3–100 Hz), red (100–250 Hz).

Nuclei of epithelial cells are clearly visible in Fig. 5b,e and in magnified view in Fig. 5g that also show structures consistent with microvilli on the epithelial cells. In addition to enterocytes, other cell types, likely goblet cells, are visible in Fig. 5e and in magnified view in Fig. 5h. Highly active cells within the villi, located in the lamina propria, are shown in Fig. 5c and in magnified view in Fig. 5i. These cells are likely antigen-presenting cells, lymphocytes, and mesenchymal elements of the intestinal lamina propria, such as myofibroblasts, fibroblasts, and mural cells[31]. Vascular networks can also be identified in Fig. 5f.

**Small Intestine Imaging (serosal side):** Figure 6 presents images of the mouse small intestine acquired from the serosal side. The myenteric and submucosal plexuses[32] are visible in Fig. 6a-b, respectively, displaying neurons and glial cells. The submucosal plexus lies embedded in the connective tissue of the submucosa and typically runs alongside blood vessels, as shown in Figures 6b–d.

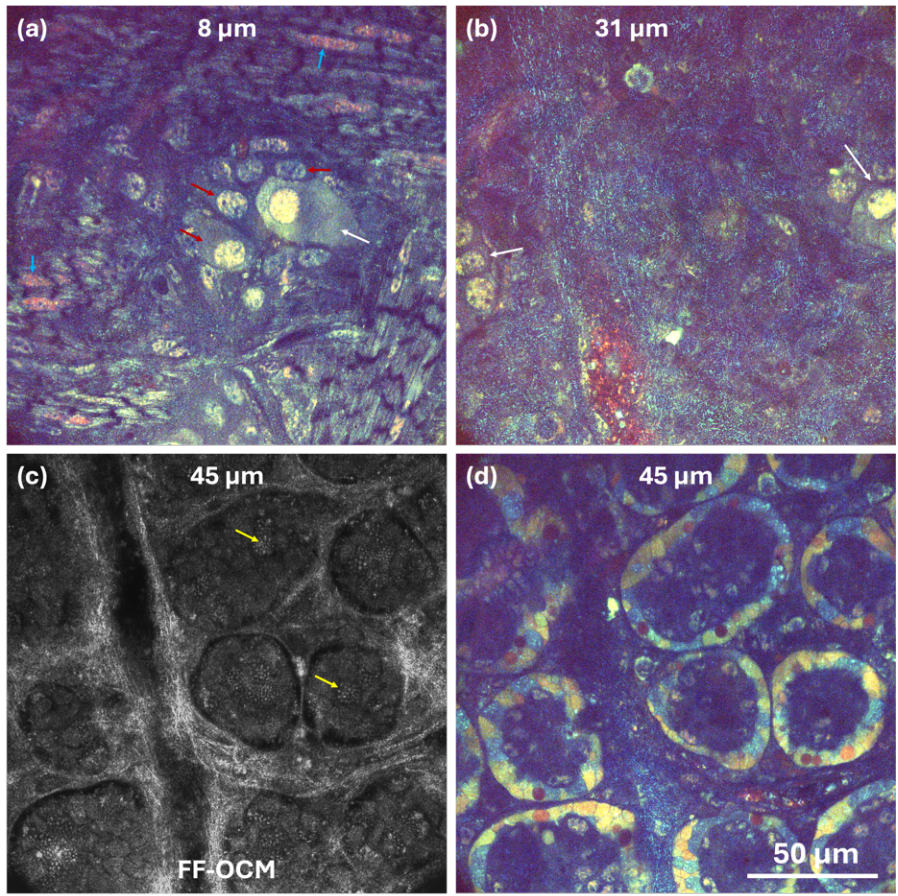

Fig. 6. FF-OCM of mouse gut from the serosal side reveals enteric plexuses and crypt architecture (taken from Supplementary Video 5). (a) *d*-FF-OCM of the myenteric plexus showing neurons (white arrow), glial cells (red arrows), and adjacent longitudinal and circular muscle fibers with visible nuclei (blue arrow). (b) *d*-FF-OCM of the submucosal plexus highlighting neuronal and glial cell bodies at the image edges (white arrows). (c) Static FF-OCM of intestinal crypts visualizing Paneth cells and their characteristic granules (yellow cells). (d) *d*-FF-OCM of the same crypt region revealing epithelial cells and underlying submucosal blood vessels. Dynamic fluctuations are color-coded by frequency: blue (~1 Hz), green (2–50 Hz), red (50–250 Hz).

Moreover, since the myenteric plexus is already visible between the longitudinal and circular muscle layers (Fig. 6a and Supplementary Video 5), the deeper neuronal network must correspond to the submucosal plexus. Supplementary Video 5 depicts additional layers of both plexuses. This dataset is unique in capturing both networks in the same z-stack. Figure 6c highlights Paneth cells, specialized epithelial cells of the small intestine[33], located at the base of intestinal crypts. These cells are identifiable in standard FF-OCM images (Fig. 6c) by their readily discernible secretory granules, characteristic of Paneth cells, which are the only granule-containing cell type at the crypt base. The Paneth cells and their granules were even more visible in fixed mouse gut, as shown in Supplementary Fig. S2 and Supplementary Video 6. Figure 6d shows *d*-FF-OCM image of round, ring-shaped profiles (the crypt lumens) surrounded by a band of brightly colored cells—crypt epithelial cells. Figures 6c and 6d provide complementary information - structural and metabolic insights into the crypts, respectively. Spindle-shaped inter-crypt cells that are subepithelial myofibroblasts (and fibroblasts) can also be identified in Supplementary Video 5.

## 4. Discussion and conclusions

We achieved 270 nm lateral and 500 nm axial resolution (Fig. S1) using 100× oil-immersion objectives and a laser-pumped white light source. The high NA objective also enabled more efficient light collection from a sample, which together with the high source brightness enabled collecting more photons and increasing SNR. Furthermore, real-time reference arm adjustment ensured coherent and confocal gate overlap, producing high contrast images up to 100 μm in fresh *ex vivo* mouse tissues, corresponding to 5–6 hepatocyte layers (Fig. S3). Moreover, the source enabled 500 fps acquisition instead of the typical 100 fps in *d*-FF-OCM. This faster imaging rate allowed frequency analysis extension from up to 30-50 Hz[6,19,23,24,34] to up to 250 Hz. This extension allowed the capture of faster dynamic processes and generation of more contrasted *d*-FF-OCM images, as illustrated in Fig. S4. For example, some of the structures with 0-30 Hz analysis in Fig. S4(e) are not well separated but become separated with 0-250 Hz analysis, shown in Fig. S4f. Also, as shown in Fig. S4g, frequency spectrum extends well above the usual 30-50 Hz range. While RGB segmentation (blue: low ~1–2 Hz; green: mid 3–100 Hz; red: high 100–250 Hz) offers intuitive visualization, it relies on arbitrary binning and may not fully capture spectral essence. Alternatives like HSV mapping, autocorrelation decay, or singular value decomposition (SVD)[35] could provide more comprehensive representations, though potentially at the expense of interpretability. These improvements in resolution and vibration spectrum analysis allowed us to better resolve filamentous networks in hepatocytes (likely related to mitochondrial dynamics[19]) and heterogeneous nuclear activity (Figs. 3–4) in liver. Sinusoids displayed broadband, high-frequency signals, enabling visualization of erythrocytes and putative immune cells (e.g., Kupffer or NK cells, based on size, morphology, and location) at depths up to 100 μm (Fig. S3), which is in contrast to previous *d*-FF-OCM[19,36] and scanning dynamic OCT[37-41] studies on liver that do not clearly visualize sinusoids or distinctly resolve individual hepatocyte layers, likely due to limited SNR or spatial resolution. In small intestine, we captured microvilli, goblet cells, and lamina propria elements from the mucosal side (Fig. 5), comparable to two-photon autofluorescence imaging of metabolic activity through detection of NADH and FAD signal[42,43]. Compared to *d*-FF-OCM on human mucosa[44], our results show superior subcellular detail in mouse villi and crypts. From the serosal side, we report the first OCT images of the submucosal plexus (Fig. 6b, Supplementary Video 5) and Paneth cells (Fig. 6c, S2, Supplementary Video 6), complementing prior lower-resolution views of the myenteric plexus using FF-OCM[45] and *d*-FF-OCM[46]. A comparison between static and dynamic images acquired at the crypt base reveals cellular populations distinct from Paneth cells. Given their specific anatomical localization within the crypt, these additional cells are hypothesized to be stem cells[47]. The epithelial cells within the crypts are clearly discernible (Fig. 6d). Future studies should aim to identify their specific types, which may exhibit heterogeneity, using correlative imaging. Although dual FF-OCM and fluorescence microscopy systems can be readily developed[48], the integration of *d*-FF-OCM with two-photon microscopy[49,50] would be most valuable. This is due to two-photon microscopy's superior depth penetration with intrinsic optical sectioning and reduced phototoxicity, providing a powerful complement to *d*-FF-OCM dynamic contrast. Further frequency expansion beyond 250 Hz could be achieved by utilizing a faster camera, as higher-frequency components of cellular dynamics may lie within the vibrations of structures such as nuclei and other organelles. While the system could be further improved with even higher NA objectives, such as, for example, NA=1.45, and a brighter, spectrally broader light source, it is already operating near optimal parameters. Increasing NA would require more complex objective lenses, potentially complicating chromatic dispersion matching and raising costs significantly (from ~€1,500 for current objective to >€10,000 for the highest NA one). Expanding spectrum to, for example, 400 nm would exacerbate the dispersion mismatch, and moreover, introduce more phototoxic light to samples. Similarly, additionally increasing intensity on a sample—via higher power or brightness of the source—would raise phototoxicity risk even without more harmful 400-500 nm spectral part. Although tissues like liver tolerate higher powers than isolated cell lines, thanks to extracellular matrix support, superior heat dissipation and the absence of labeling, further increases in power could prove counterproductive. Nonetheless, the use of 50 mW of illumination power over a 173 μm × 173 μm field of view—corresponding to an irradiance of 170 W/cm²—may be acceptable for imaging a variety of tissues. For comparison, continuous exposure to 561 nm light at 1 kW/cm² for several minutes has been used to excite fluorescence in live NIH 3T3 cells without any observed adverse effects[51]. Moreover, imaging mitochondria—one of the main contrast sources in dynamic OCT—has been performed at irradiances as high as 10 kW/cm² at 561 nm for up to one minute before noticeable morphological changes occurred[52]. No signs of phototoxicity were observed in the reported results, despite unnecessarily prolonged light exposure during image saving, which took five times longer than the 1-second acquisition. In the next-generation system, this will be addressed by significantly reducing saving time—by a factor of five, which is readily achievable—and by turning off illumination during saving, as the light source supports triggering at rates of at least 100 kHz.

In conclusion, the reported system uses high-NA optics, a bright incoherent source, and extended frequency analysis to enable unprecedented *d*-FF-OCM imaging of deep tissue dynamics. By visualizing novel structures in mouse liver and intestine with fluorescence-like contrast, it holds promise for biomedical research and various clinical applications. Ultimately, *d*-FF-OCM is poised to play a role in advancing personalized medicine, enabling earlier and more precise diagnostics, and facilitating a deeper understanding of disease mechanisms in a dynamic, living context, particularly for applications like rapid intraoperative pathology where its label-free, high-resolution capabilities are invaluable.

**Funding.** Research Council of Lithuania (LMTLT), agreement No. P-MIP-22-346.

**Disclosures.** The authors declare no conflicts of interest.

**Acknowledgments.** We thank Seemantini Nadkarni and Arvydas Laurinavičius for discussions, and Urtė Neniškytė, Viktorija Kralikienė and Igor Nagula for providing mouse tissues.

**Contributions.** E.A. conceived the idea, designed the imaging system, processed the data, analyzed the results and wrote the manuscript draft. E.T., A.T and E.A. built the system and performed experiments. E.T. and E.A. interpreted the results and prepared the figures. E.T and A.T prepared tissue for imaging. All authors were involved in discussions and contributed to the manuscript editing.

**Data Availability Statement (DAS).** The data that support the findings of this study are available from the corresponding author upon reasonable request.

## 5. Supplementary

## Resolution

The theoretical lateral resolution of the FF-OCM system is diffraction-limited to approximately $\Delta x = 268$ nm at the central wavelength, $\lambda$ of 550 nm, calculated using the Rayleigh criterion ($\Delta x = 0.61\lambda/NA$) with NA=1.25. With a total magnification of 100× and camera pixel size of 12 μm, the effective pixel size in sample plane is 120 nm, supporting minimal Nyquist sampling at ~240 nm (strictly requiring at least 2 pixels per resolvable feature to avoid aliasing) and confirming the system is near the optical limit without significant aliasing.

Axial resolution was determined by measuring the full-width-half-maximum (FWHM) of an axial profile obtained from a mirror sample, like shown in Supplementary Fig. S1(b).

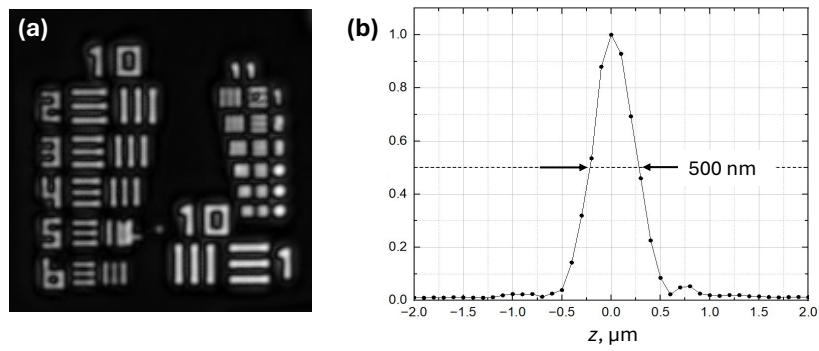

Fig. S1. Resolution Characterization of the Full-Field Optical Coherence Microscopy (FF-OCM) System. (a) Lateral spatial resolution determined with a USAF 1951 test target, where the smallest fully resolvable features correspond to 274 nm bar spacing (element 6, group 10). (b) Axial resolution measurement.

## Visualizing Paneth Cells

Paneth cells and their granules are clearly visible in fixed mouse tissue (Supplementary Fig. S2a-b), with greater clarity than in fresh *ex vivo* tissue (Fig. 6c). See Supplementary Video 6 for the full movie.

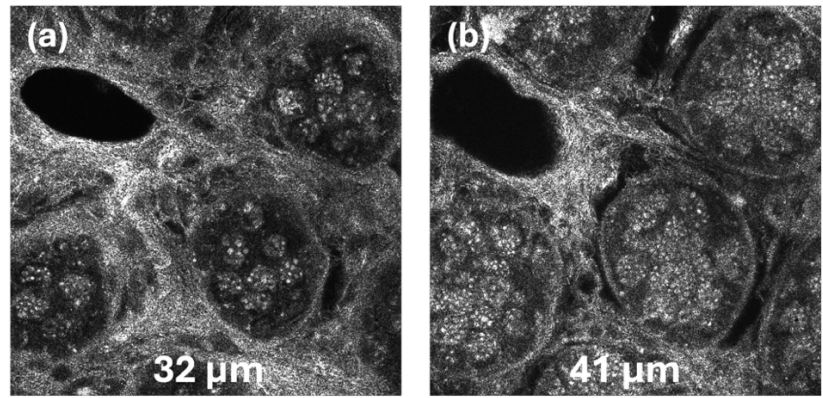

Fig. S2. Paneth cells featuring granules imaged with static FF-OCM on fixed mouse gut at different depth (a) 32 μm and (b) 41 μm.

# Demonstration of imaging depth over 100 µm

Imaging depth in liver tissue with the developed *d*-FF-OCM system can exceed 100 µm, as shown in the Supplementary Fig. S3.

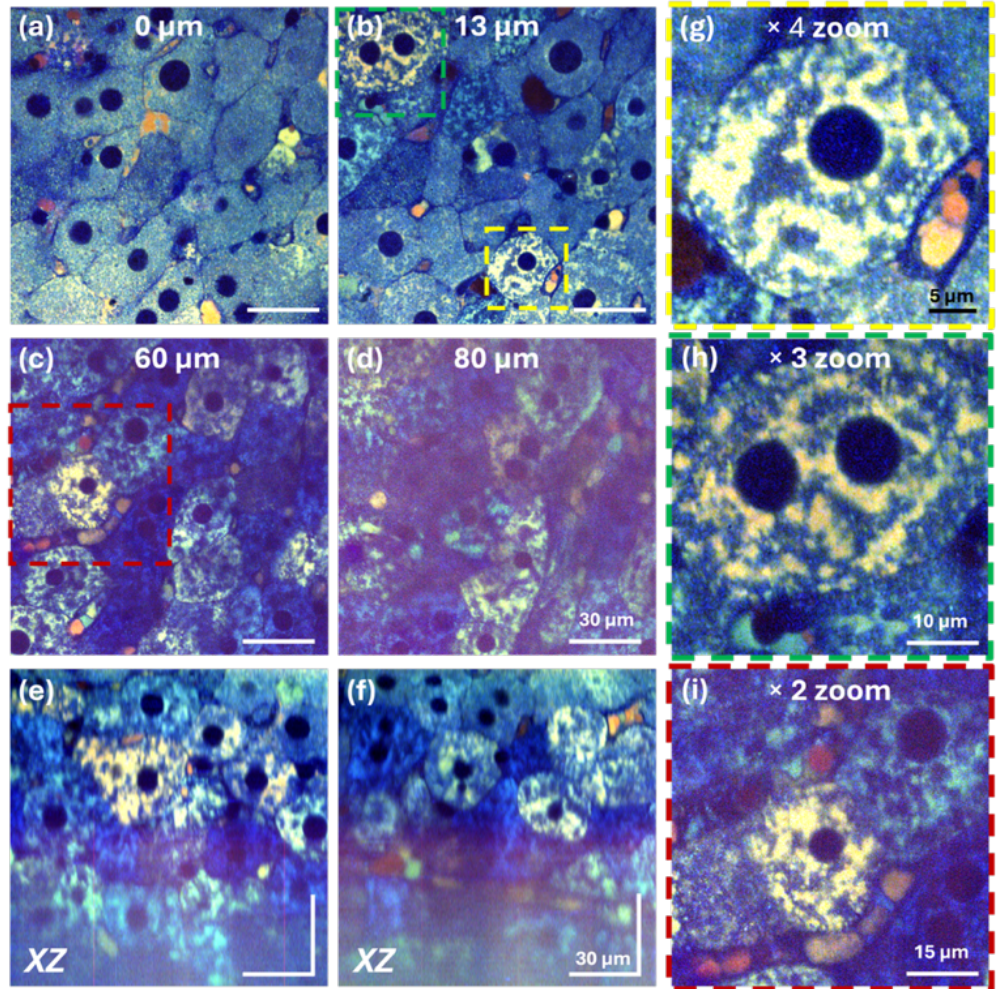

Fig. S3. Dynamic FF-OCM enables high-contrast en face imaging of freshly excised mouse liver down to 100 µm depth. Panels (a–d) show en face views at 0 µm, 13 µm, 60 µm, and 80 µm depths—note the preserved contrast even at 80 µm. Panels (e) and (f) present the full axial (*XZ*) span with 5-6 layers of hepatocytes that were derived by averaging 20 consecutive *XZ* images (B-scans). Panels (g–i) are 4× magnified detail views highlighting hepatocyte plates and blood-filled sinusoids. Dynamic vibration frequencies are color-coded as follows: blue (0.5–1 Hz), green (1.5–50 Hz), and red (50–150 Hz).

# Frequency extension to 250 Hz

Extension of frequency analysis band from the upper cut-off frequency of 30 Hz to 250 Hz yields more contrasted images, as shown in Supplementary Fig. S4.

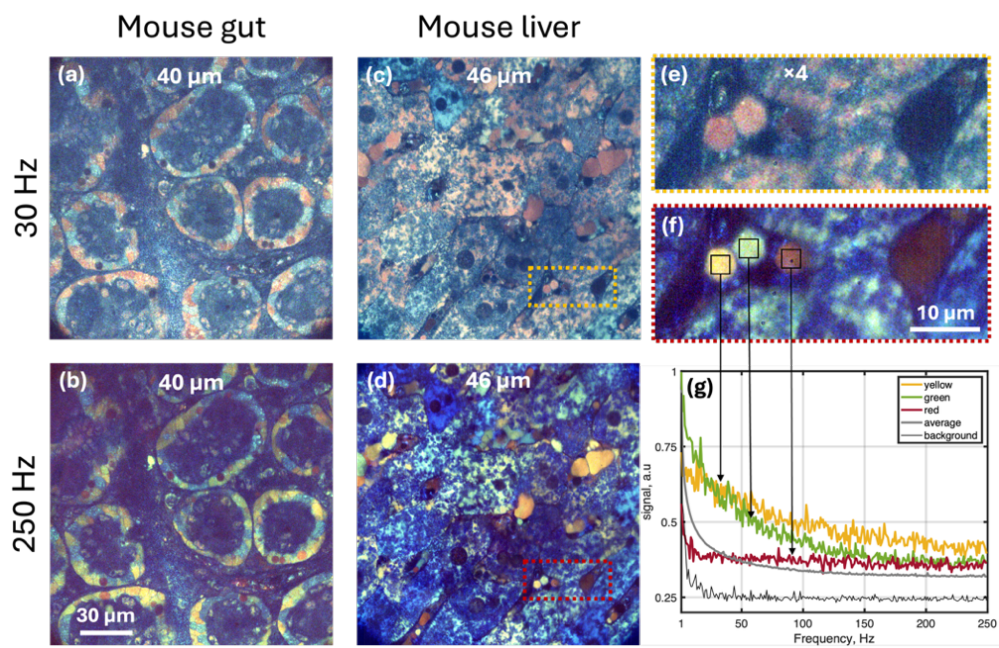

Fig. S4. Raising the upper frequency limit from 30 Hz to 250 Hz significantly improves contrast in d-FF-OCM images. Panels (a) and (b) show mouse gut images acquired with 30 Hz and 250 Hz cut-offs, respectively, and panels (c) and (d) show the corresponding liver images. Panels (e) and (f) are 4× magnified views of the liver regions highlighted by the yellow and red boxes in (c) and (d), respectively. Panel (g) plots the frequency spectra of three structures, 4-5 µm in diameter, within the liver sinusoids (f). Average and background frequency spectra are also shown. Note that dark red/brown structures visible in (f) are absent in the 30 Hz image (e) because its high-frequency components are filtered out by the 30 Hz analysis.

Supplementary videos:

**Video 1**: 3D volumetric rendering of ex vivo mouse liver acquired by dynamic FF-OCM, visualizing hepatic architecture and subcellular dynamics through a 100 µm depth stack.
**Video 2**: Depth-resolved *en face* sequence of *ex vivo* mouse liver captured by dynamic FF-OCM, illustrating tissue architecture and subcellular dynamics across multiple z-planes.
**Video 3**: Depth-resolved *en face* sequence of *ex vivo* murine small intestine (sample #1), captured from the mucosal side by dynamic FF-OCM, illustrating epithelial and lamina propria structures across multiple z-planes.
**Video 4**: Depth-resolved *en face* sequence of *ex vivo* murine small intestine (sample #2), captured from the mucosal side by dynamic FF-OCM, illustrating epithelial and lamina propria structures across multiple z-planes.
**Video 5**: Depth-resolved *en face* sequence of *ex vivo* murine small intestine, captured from the serosal side by dynamic FF-OCM, illustrating longitudinal and circular muscle, myenteric and submucosal plexuses and crypts with Paneth cells.
**Video 6**: Depth-resolved *en face* sequence of *ex vivo* murine small intestine, captured from the serosal side by static FF-OCM, illustrating longitudinal and circular muscle, myenteric plexus and Paneth cells with clear granule structure.